# Real-time optimizing of thermoelectric coolers performance based on energy and exergy analysis

Reza Jamali[1], S. Aria Hosseini[2], Alireza Amiri-Margavi[3], Farschad Torabi[4]*

[1] Mechanical Engineering Department, Amirkabir University of Technology, Tehran, Iran

[2] Mechanical Engineering Department, UC Riverside, Riverside CA 92521, USA

[3] Department of Mechanical Engineering and Materials Science, University of Pittsburgh, Pittsburgh, PA 15213, USA

[4] Department of Energy Systems, Mechanical Engineering Faculty, K.N. Toosi University of Technology, Tehran, Iran

* Corresponding author: ala170@pitt.edu

## ABSTRACT

New strategy is presented to optimize the performance of Thermoelectric (TE) coolers. This approach breaks optimizing TE coolers free from traditional methods of controlling temperature or engineering materials and the structural properties of the junctions. We introduced a dimensionless figure, γ, that shows the ratio of the unavailable cooling capacity to the available cooling capacity. This parameter relates the TE coolers coefficient of performance (COP) to the COP of the reversible cycle (second law of thermodynamics efficiency) for a given electrical current. The theoretical description of the model is presented, and it is shown that controlling γ during the TE performance minimizes entropy generation and energy loss, which leads to the maximum pumped heat. We validated this model against a designed TE cooler. In this cooler, contrary to conventional TE coolers, where the temperature of the cold space is generally controlled at a specific temperature, and the performance of the cooler overlooked, the entropy generation and heat loss are engineered, and the electrical current is tuned to minimize γ by the controller so that the TE cooler works near to its optimum performance at any time.

**Nomenclature**

| | |
|---|---|
| A | Seebeck coefficient (VK$^{-1}$) |
| I | Electrical current (A) |
| k | Thermoelectric thermal conductance (WK$^{-1}$) |
| l | Efficient length (m) |
| L | Heat exchanger thermal conductance (WK$^{-1}$) |
| N | Number of n-p doped |
| Q | Heat flow rate (W) |
| R | Electrical resistance (Ω) |
| s | Entropy |
| S | Surface area (m$^2$) |
| T | Temperature (K) |
| V | Voltage (V) |
| W | Electric energy consumption thermoelectric materials |
| ZT | figure of merit |

**Greek letters**

| | |
|---|---|
| $\gamma$ | Objective function |
| $\Delta$ | Differential |
| $\rho$ | Electrical resistivity (Ωm-1) |
| $\eta$ | Efficiency of the second law of thermodynamic |
| $\alpha$ | Seebeck coefficient (VK$^{-1}$) |

**Scripts**

| | |
|---|---|
| C | Cold side of the thermoelectric cooler |
| Cj | Junction temperature in the cold side of thermoelectric |
| Hj | Junction temperature in the hot side of thermoelectric |
| max | Maximum available value |
| loss | lost energy |
| gen | Generation |
| i | Index |
| n | n type material |
| p | p type material |
| reversible | reversible thermodynamic cycle |
| surr | Surrounding |
| sys | System |

**Abbreviation**

| | |
|---|---|
| COP | Coefficient of performance |
| TE | Thermoelectric |
| TEC | Thermoelectric Cooler |

# INTRODUCTION

Thermoelectric (TE) coolers use the Peltier effect to directly convert electricity into heat. These coolers show numerous benefits, including simple structural design with no moving parts, which lower the risk of failure due to fatigue and fracture [1-4]. Current controlled semiconductor-based components which help to control the heat harvesting by tuning the electrical current magnitude and direction [5]. These coolers are compact, lightweight, and noise-free solid-state devices. TE coolers are environmental friendly and do not use refrigerant gases in their operation. Unfortunately, these coolers demonstrate lower energy efficiency compare to the mechanical counterparts that hindered their expansion to real applications[6]. However, theoretical predictions suggested that there is a huge room to enhance the energy efficiency of these materials, which led to a wide range of theoretical and experimental efforts to improve the performance of these materials, lower the wasted cooling capacities, and optimize heat exchanger thermal performance [7-11].

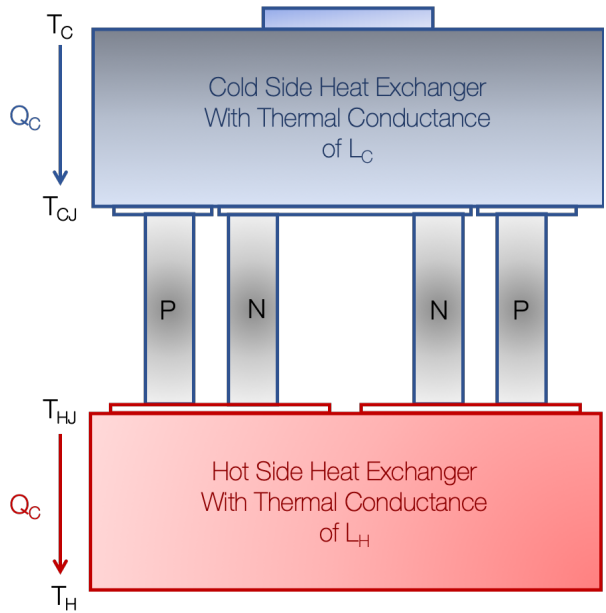

**Figure 1:** Schematic view of TE cooler. Heat travels from the cold side toward the hot side. Cold side, hot side and the PN junctions behave as thermal resistance.

The first studies on optimizing *TE* coolers using the entropy flow approach were reported in 90s developed a phenomenological model to analyze the behavior of *TE* coolers to capture the interface effect of *TE* coolers [12, 13]. The temperature-entropy approach is used to calculate the useful cooling capacity and energy consumed by a dissipative effect to demonstrate the performance of *TE* coolers [14-17]. A numerical optimization approach proposed to design an optimum *TE* cooler [18]. They used geometry configuration, electrical current, number of *TE* junctions, and the cold side temperature of the *TE* coolers as independent variables. They maximize the total cooling rate for different operating conditions in the presence of limitations on the electrical power supply. A study by Solkbrakken et al. [19] explored an electrical current that minimized the junction temperature in the existence of a thermal resistance heat sink. Later research [20] showed that it is possible to choose an operating current for *TE* coolers that both maximizes the COP and minimizes the junction temperature. Taylor et al. [21] demonstrated an optimization method for *TE* coolers that are used in electronic devices. In their study, COP maximization and junction minimization approaches for small heat sinks are used. The independent variables were electrical current and *TE* coolers' geometry. For a given heat load and heat sink thermal resistance, they showed that the electrical current applied to the TEC module and TEC geometry can be tuned to minimize the junction temperature and maximize the COP. Cran and Jackson [22] conducted a study on the effect of cross-flow heat exchangers coupled to a *TE* cooler with the ratio of the effective thermal power to the cost of the *TE* cooler as the objective function. Machine learning approaches can be employed to optimize the performance of *TE* coolers and enhance their efficiency [23-25]. David et al. [26] proposed an optimization method based on coupling the thermoelectric phenomena to heat transfer and pressure drop in the heat exchangers. The objective function was the maximizing of the COP, therefore, the flow rate of water on the heat exchanger; and the *TE* geometry were the variables that should be obtained for the aim of optimizing the objective function. Their results agreed well with the results obtained from entropy generation minimization method. Yang et al. [27] proposed a multi-objective optimization for *TE* cooler in order to make a proper balance between surface temperature, operating current, and hot-side thermal resistance. In some cases, their optimized variables were not matched with real application. They concluded that more parameters like airflow rate, cold-side thermal resistance should be considered to increase the accuracy of the optimal values. Recently, Lu at el. [28], proposed a Multi-objective optimization for *TE* cooler. They analyzed the influence of applied current, geometry configuration, contact resistance, and electrical contact resistance on *TE* cooler through three-dimensional numerical simulations. They optimized two objective functions, cooling capacity and COP using genetic algorithm. Eventually, they proposed a useful procedure for designing *TE* coolers. For a recent review on the optimization, mathematical models and machine learning models for optimization references [29-37] provide a myriad of models and recent reviews.

The performance of *TE* coolers relies on the intrinsic properties of *TE* materials, heat exchanger capability, and working conditions. The working conditions, including temperature gradient between hot and cold sides, input current and voltage, and properties of the coolers may change over time; therefore, developing design strategies of self-adaptive *TE* coolers that can optimize the performance in real-time is inevitable. In this paper, we present a new strategy for optimizing *TE* coolers that breaks *TE* coolers' design free from traditional methods of controlling temperature or engineering materials and the structural properties of the junctions. The theoretical description of entropy generations and heat loss capacity is presented, and an objective function is introduced. The central idea of this objective function is to relate cooling capacity loss to actual cooling capacity. This parameter is directly related to the second law of thermodynamics efficiency. It is shown how minimizing the parameter leads to the optimum performance of the *TE* coolers. A *TE* cooler is designed based on the developed theory and is validated against theoretical predictions.

## ENERGY ANALYSIS OF *TE* COOLERS

Thermoelectric cooler modules are made up of a series of n-doped and p-doped semiconductors. The performance of thermoelectrics is widely described by the dimensionless figure of merits, $ZT = \frac{A^2}{KR}$, where A is the Seebeck coefficient, R is electrical resistance, and K is the thermal conductance. For a thermoelectric with N n-doped and p-doped channels, A, R and K define as

$$R = (\rho_p + \rho_n)\frac{1}{S}N, \tag{1}$$

$$A = (\alpha_p + \alpha_n)N, \tag{2}$$

$$K = (\kappa_p + \kappa_n)\frac{S}{l}N, \tag{3}$$

where, l and S are the effective length and the Surface area of the channels, respectively [7]. The energy balance equations for the hot and cold sides lead to

$$AIT_{Hj} - \frac{1}{2}RI^2 - K(T_{Hj} - T_{Cj}) - Q_C = 0, \tag{4}$$

$$AIT_{Hj} - \frac{1}{2}RI^2 - K(T_{Hj} - T_{Cj}) - Q_H = 0, \tag{5}$$

where $Q_C$ is the heat flow through the cold side heat exchanger, $Q_H$ is the heat flow through the hot side heat exchanger, I is the electrical current and $T_{Cj}$ and $T_{Hj}$ are the junction's temperature for the cold and hot sides, respectively [33, 34]. For a given environmental condition, *TE* cooler's performance only relies on electrical current, I. The heat flow in the cold and hot sides of the *TE* coolers can be written as functions of electrical current (Q = Q(I)). The heat balance for heat exchangers describes as

$$Q_C = L_C(T_C - T_{Cj}), \tag{6}$$

$$Q_H = L_H(T_{Hj} - T_H), \tag{7}$$

where $T_C$ and $T_H$ are cold and hot space temperature, respectively, $L_C$ and $L_H$ are cold and hot heat exchanger thermal conductance. The $T_{Cj}$ – independent and $T_{Hj}$ – independent forms of equations (4) and (5) derive by substituting $T_C$ and $T_H$ from equations (6) and (7).

$$AI\left(T_c - \frac{Q_c}{L_c}\right) - \frac{1}{2}RI^2 - K\left[\left(T_H + \frac{Q_H}{L_H}\right) - \left(T_c - \frac{Q_c}{L_c}\right)\right] - Q_c = 0, \tag{8}$$

$$AI\left(T_H + \frac{Q_H}{L_H}\right) + \frac{1}{2}RI^2 - K\left[\left(T_H + \frac{Q_H}{L_H}\right) - \left(T_c - \frac{Q_c}{L_c}\right)\right] - Q_H = 0. \tag{9}$$

Equations (4) and (5) are a system of linear equations that can be solved for $Q_c$ and $Q_H$. The final forms of the heat kernels are

$$Q_C = L_C \frac{AIT_C - \frac{1}{2}RI^2 - K(T_H - T_C) - K\frac{Q_H}{L_H}}{AI + K + L_C}, \tag{10}$$

$$Q_H = -\frac{AIT_H - \frac{1}{2}RI^2 - K(T_H - T_C) - K\frac{AIT_C - \frac{1}{2}RI^2 - K(T_H - T_C)}{AI + K + L_C}}{-\frac{AI}{L_H} + \frac{K^2}{L_H(AI + K + L_C)} - \frac{K}{L_H} - 1}. \tag{11}$$

For a given surrounding condition, equations (10) and (11) only depend on the electrical current which is controlled by the controller. Similarly, *TE* energy consumption, W(I), *TE* voltage, V(I), and coefficient of performance, COP(I), obtain from the magnitude of $Q_C$ and $Q_H$ as

$$W(I) = Q_H(I) - Q_C(I), \tag{12}$$

$$V(I) = \frac{Q_H(I) - Q_C(I)}{I}, \tag{13}$$

$$COP(I) = \frac{Q_C(I)}{Q_H(I) - Q_C(I)}. \tag{14}$$

## OPTIMIZATION TARGET FUNCTION

Unfortunately, electrical current produces countervailing responses in COP, W and $Q_C$. These parameters are shown for TEC1-12704 thermoelectric cooler with the constant surrounding condition of $T_C = 300K$, $T_H = 305K$, $L_C = 1\frac{W}{K}$, $L_H = 2\frac{W}{K}$ in figure 2a. Although both W and $Q_C$ increase by increasing the electrical current, W is more sensitive to the changes in electrical current, therefore the denominator in equation (14) changes faster than the numerator, thus COP decreases by increasing the electrical current. figure 2b shows a strong dependency of COP, $Q_H$ and $Q_H$ on the temperature gradient between the thermoelectric junctions. For a large temperature gradient, both COP and $Q_C$ lower significantly.

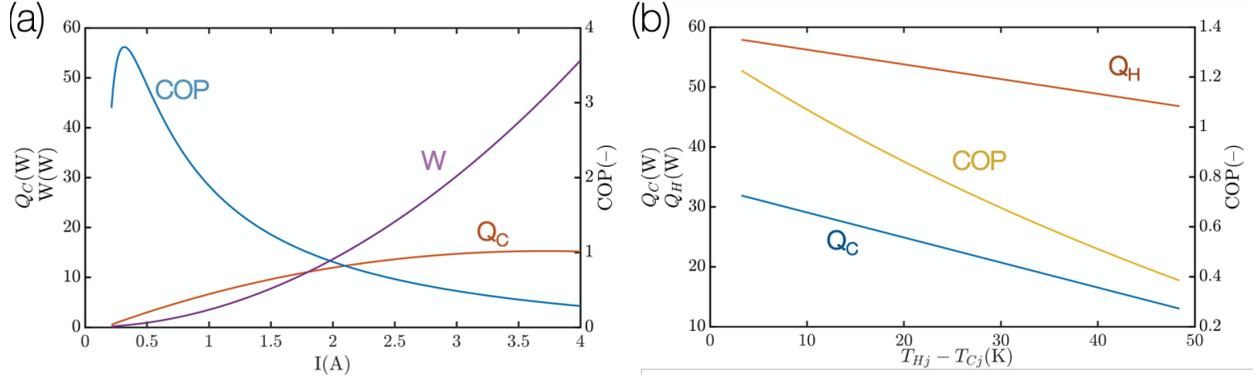

**Figure 2:** TE parameters of TEC1-12704: (a) The energy consumption is growing faster than available cooling capacity. So, the denominator of equation (14) increases more than the numerator, and coefficient of performance decreases. We conclude that COP is not an appropriate optimization object. (b) High temperature gradient between hot and cold junctions leads to low COP and cooling capacity.

Figure 3 shows the coefficient of performance (COP) of TEC1-12704 *TE* cooler for different electrical current, I, and temperature gradient. The maximum COP happens at a very low electrical current that returns near zero refrigeration load; Therefore, the magnitude of COP is not an appropriate function for optimizing the cooling capacity in this case. An alternative strategy is to minimize the ratio of unavailable cooling capacity to available cooling capacity. We introduce a dimensionless figure, γ, as

$$\gamma = \frac{Q_{C_{loss}}(I)}{Q_C(I)}. \quad (15)$$

This dimensionless parameter shows the $Q_{C_{loss}}(I)$ normalized by the actual cooling capacity. Our ultimate goal is to minimize the irreversibility which leads to the maximum amount of cooling capacity, $Q_C$, and minimum unavailable cooling capacity, $Q_{C_{loss}}$. The parameter γ is inversely proportional to the amount of pumped heat and directly proportional to the amount of unavailable cooling capacity; Therefore, we have chosen γ as the target function for optimizing. Using equations (12 − 14) and (A.5)

$$\gamma = \eta_{II} - 1. \tag{16}$$

where $\eta_{II}$ is the second law of thermodynamics efficiency describes as

$$\eta_{II} = \frac{COP_{reversible}(I)}{COP_{TEC}(I)}. \tag{17}$$

In this equation, $COP_{reversible} = \frac{T_{Cj}}{T_{Hj}-T_{Cj}}$ shows the performance of the ideal reversible cycle of Carnot.

Figure 4 shows γ(I) for different surrounding conditions for *TE* cooling module model TEC1-12704. In figure 4a γ(I) for $T_c = 300K$, $L_c = 1W/K$, $L_H = 2W/K$, and different values of $T_H$ is plotted. Figure 4b shows γ(I) for $T_H = 305K$, $L_c = 1\frac{W}{K}$, $L_H = 2\frac{W}{K}$, and different values of $T_c$. The dashed line in all cases shows the optimum electric current that minimizes γ for different environmental conditions. This optimum electrical current which minimizes the objective function of γ varies by time. The adaptive designed controller should calculate this optimum electrical current in real-time and apply it to the *TE* cooler. Therefore, the *TE* cooler works at its optimum state during the performance. Another important environmental parameter is the thermal conductance of the heat exchanger. Air is commonly used as the cooling bath. Changes in the rate of airflow and thermal properties of the air may change the thermal conductance of the heat exchangers. The adaptive controller system can optimize the performance of *TE* coolers by tuning the electrical current which minimizes γ according to the thermal properties of the heat exchangers. Figure 4c shows γ(I) for $T_c = 300K$, $T_H = 330K$ and different values of thermal conductance of the hot side heat exchanger. In figure 4d, the same surrounding conditions are considered except that thermal conductance of the cold side heat exchanger that is changed. The optimum electrical current leads to the best performance of the *TE* coolers strongly depend on the surrounding conditions except for the thermal conductance of the cold side heat exchanger as shown in figure 4d.

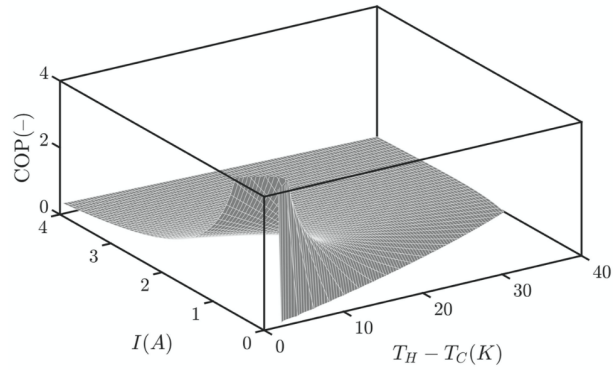

**Figure 3:** The coefficient of performance for TEC-12704 as a function of electrical current and temperature gradient. The maximum COP happens at very low electrical current.

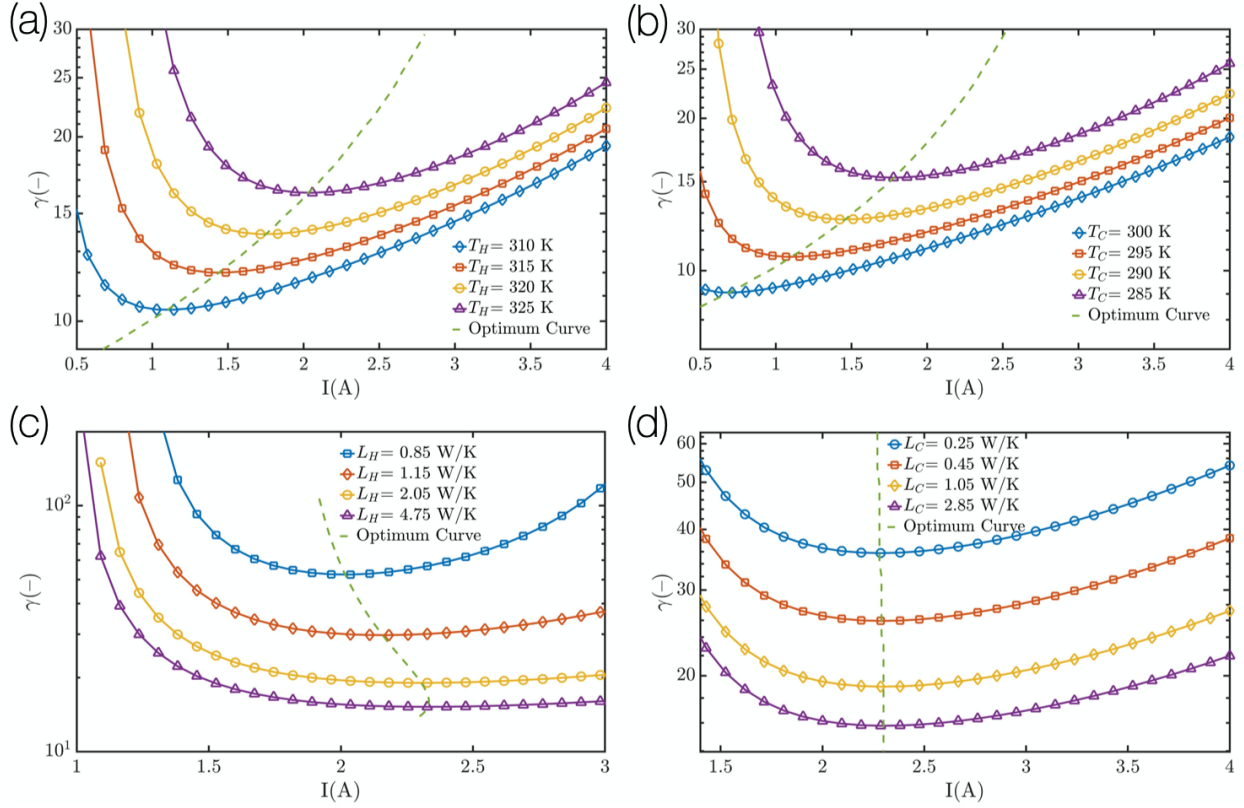

**Figure 4:** Objective function in TEC-12704 for different environmental parameters, the dash line in all cases shows the optimum electrical current. (a) In this plot $T_c = 300K$, $L_c = 1W/K$, $L_H = 2W/K$, and $T_H$ changes from $310\ K$ to $325\ K$. The objective function and the optimum electrical current increase by increasing $T_H$. (b) This plot shows $\gamma(I)$ for $T_H = 305K$, $L_c = 1\frac{W}{K}$, $L_H = 2\frac{W}{K}$ and $T_c$ varies from $285\ K$ to $300\ K$. Contrary to previous plot, the objective function and the optimum electrical current decrease by increasing $T_c$. (c) $\gamma(I)$ for $T_c = 300K$, $T_H = 330K$ and different values of thermal conductance of the hot side heat exchanger from $0.85\frac{W}{K}$ to $4.75\frac{W}{K}$ is plotted. Increasing $L_H$ lower $\gamma(I)$. The optimum electrical current increases by increasing $L_H$. (d) Same surrounding parameters as plot (c) is considered except for cold side heat exchanger. $\gamma(I)$ decrease by increasing $L_C$ but the optimum electrical current is insensitive to $L_C$.

## EXPERIMENTAL SETUP

Figure (5) shows the components of the designed *TE* cooler. We used two *TE* cooler modules of TEC1-12704. These modules are 4 by 4 cm² in size and are used as heat transfer plates between cold and hot sides. The applied voltage and the current vary between 0-12 (V) and 0-4 (A), respectively. A set of fins are used as the heat sinks to dissipate heat to the environment. The heat transfer between the sinks and *TE* coolers are eased using silicon paste. Four temperature sensors measure the temperature of *TE* coolers' sides, inlet and outlet of the heat sinks. Other components include a centrifugal fan to create air convection, a printed circuit board with microcontroller and power supply circuit, Silicon paste, battery and connection lines. The air is injected by radial fans. The ratio of the airflow toward the hot side is twice the amount of air flows toward the cold side. Sensors record the information, including the environmental temperature, the temperature at cold and hot sides of the heat exchanger, applied voltage, and electrical current. Microcontrollers process and calculate the electrical current that minimizes $\gamma(I)$. Therefore, the *TE* cooler performs at the optimum condition in real-time. Figure 5a shows the Infrared thermograph imaging during this test. The dark area shows the cold space, and the yellow color shows the warm zone.

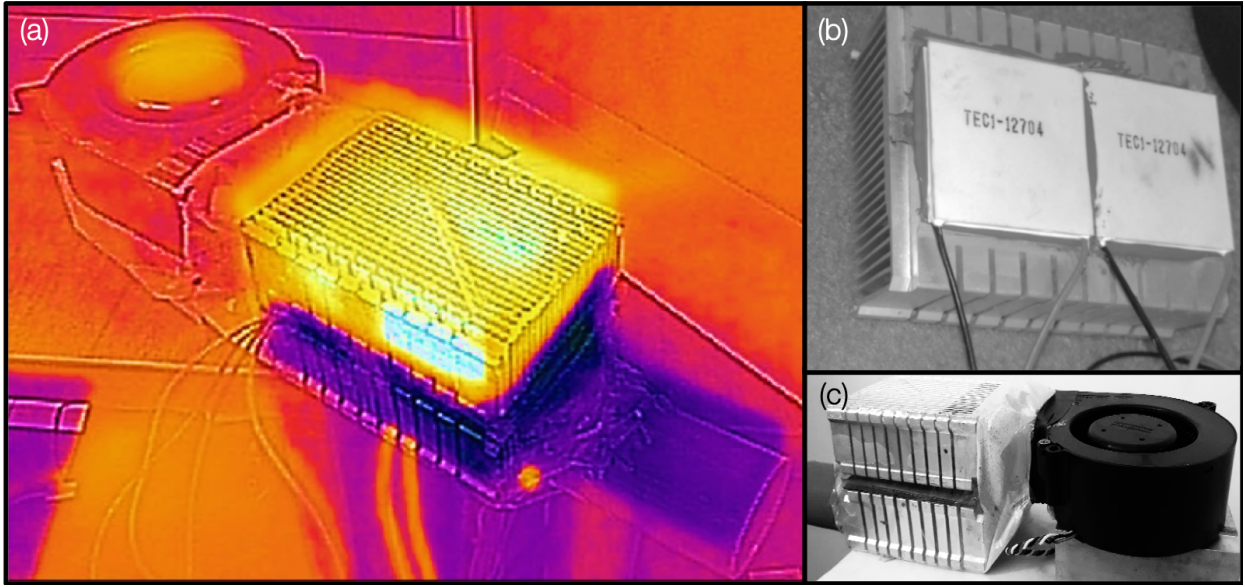

**Figure 5:** (a) Infrared thermography image of TE during its performance. The dark spots are cold area and the yellow spots are the hot area. (b-c) The component of the TE coolers and the adaptive controller used to tune electrical current to $\gamma(I)$ minimize in real-time.

$Q_H(t)$, $Q_c(t)$ and $W(t)$ are shown in figure 6a. The maximum pumped heat happens at the t=0 when two plates have the same temperature (figure 2b shows $Q_H$, $Q_c$ and COP for temperature gradient between two sides.) The heat flow is from the cold side toward the hot side of the thermo-electrics due to the Peltier effect (figure 1); therefore, it takes time for the surrounding temperatures to reach their steady-state points. The inverse heat transfer due to Peltier effects in *TE* coolers increases the temperature gradient between the cold and hot side so that the $Q_c(t)$ lowers during the time until reaching its steady-state condition. Figure 6b shows $\gamma(t)$, $\text{COP}_{\text{reversible}}(t)$, $\text{COP}_{\text{TEC}}(t)$. Both $\text{COP}_{\text{reversible}}(t)$ and $\text{COP}_{\text{TEC}}(t)$ decrease during the time however $\text{COP}_{\text{reversible}}(t)$ changes faster, therefore, $\gamma$ shows a decreasing trend during the time. Unavailable cooling capacity, actual cooling capacity, and $\gamma$ have been shown in figure 7. The lost cooling capacity decreases faster than the captured cooling capacity that also reduces $\gamma$. However, the irreversibility has been weakened during the time, therefore, the designed controller has handled its duty successfully during the experiment.

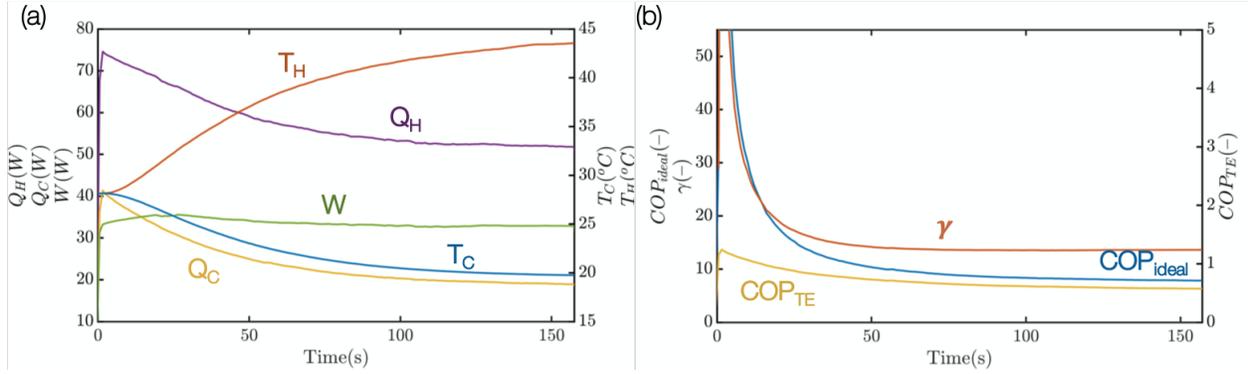

**Figure 6:** (a) TE parameters of $Q_H(t)$, $Q_c(t)$ and $W(t)$ are shown. The air temperature changed from $303\ K$ inlet air to $295\ K$ in the outlet for the cold space, and $303\ K$ inlet air to $316\ K$ for the warm space. (b) $\gamma$, $COP_{reversible}$, $COP_{TEC}$ are plotted during the TE performance. Both $COP_{reversible}(t)$ and $COP_{TEC}(t)$ decrease during the time however $COP_{reversible}(t)$ changes faster therefore $\gamma$ shows decreasing trend during the time.

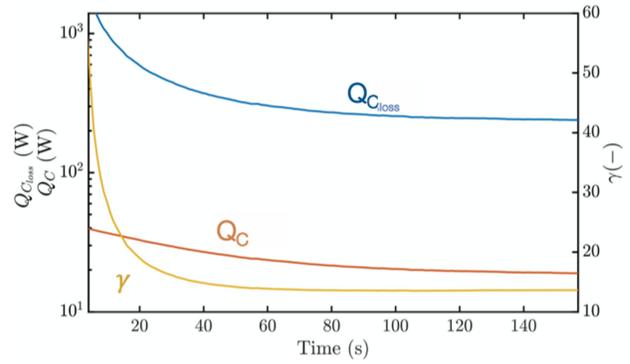

**Figure 7:** Unavailable cooling capacity, actual cooling capacity, and $\gamma$ are plotted. The lost cooling capacity decreases faster than the captured cooling capacity. The irreversibility has been weakened during the time; therefore, the designed controller has handled its duty successfully during the experiment.

## CONCLUSION

A new strategy is presented to optimize *TE* coolers. In this method, the ratio between unavailable cooling capacity to available cooling capacity is defined as the objective function. This optimization parameter is directly related to the second law of thermodynamics efficiency. The optimum electrical current that minimizes this ratio for TEC1-12704 for different environmental variables is presented. A controller designed to minimize this ratio and find the optimum electrical current to minimize the entropy generation. In fact, this study has shown that the irreversibility has been weakened during the time, and the controller could optimize the performance of the *TE* cooler.

# APPENDIX

## IRRIVESIBILITY AND ENTROPY GENERATION

The irreversibility of any thermodynamics cycle is largely guided by entropy generation during the thermodynamics process. The available energy loss is directly related to the magnitude of entropy generation. The amount of entropy generation describes as

$$s_{gen} = \Delta s_{total} = \Delta s_{sys} + \Delta s_{surr}. \tag{A.1}$$

For steady-state process $\Delta s_{sys} = 0$, therefore, $s_{gen}$ is reduced to

$$s_{gen} = \Delta s_{surr} = \sum \frac{Q_i}{T_i} = \frac{Q_H}{T_{Hj}} - \frac{Q_C}{T_{cj}}. \tag{A.2}$$

For the reversible process, we harvest the maximum amount of pumped heat. In an ideal reversible thermodynamic cycle, the entropy generation is zero ($s_{gen} = 0$), therefore

$$\frac{Q_H}{T_{Hj}} - \frac{Q_C}{T_{cj}} = 0. \tag{A.3}$$

According to the first law of thermodynamics, $Q_H = Q_C + W$. Consequently, the maximum refrigeration load for a given power consumption in an ideal reversible *TE* cooling process is

$$Q_{c_{max}} = \frac{wT_{cj}}{T_{Hj} - T_{Cj}}. \tag{A.4}$$

For irreversible cycles with energy loss, the refrigeration load would be less that $Q_{c_{max}}$. This available energy loss calculates as

$$Q_{c_{loss}} = \frac{wT_{cj}}{T_{Hj} - T_{Cj}} - Q_C. \tag{A.5}$$

The source of irreversibility in *TE* coolers can be the heat dissipation because of the temperature gradient in *TE* plates or electrical resistance in the leads and channels. Although any real thermodynamics cycle includes some entropy generation, design strategies to minimize the entropy generation should be taken for more efficient performance.